\documentclass[pre,twocolumn,amsmath,amssymb,superscriptaddress,a4paper]{revtex4}
\usepackage{natbib}

\usepackage{dcolumn}
\usepackage{graphicx}
\usepackage{bm}
\usepackage{scrtime}[24hr]
\usepackage[usenames]{color}

\usepackage{rcs} 

\RCS $Revision: 1.15 $ 
\RCSdate $Date: 2007/01/20 14:59:21 $ 

\newcommand{\degc}{\ensuremath{^{\circ}}C}

\begin{document}

\bibliographystyle{apsrev}

\title{Direct measurement of the effective charge in nonpolar suspensions by optical
tracking of single particles}

\author{G. Seth Roberts}
\affiliation{School of Chemistry, University of Bristol, Bristol BS8
1TS, UK.}
\author{Tiffany A. Wood}
\affiliation{School of Chemistry, University of Bristol, Bristol BS8
1TS, UK.}
\author{William J. Frith}
\affiliation{Unilever Research Colworth, Colworth House, Sharnbrook,
Bedford MK44 1LQ, UK.}
\author{Paul Bartlett} \email{p.bartlett@bristol.ac.uk}
\affiliation{School of Chemistry, University of Bristol, Bristol BS8
1TS, UK.}

\date{\small Revision: \RCSRevision, \RCSDate, \RCSTime\ (UTC) } 

\begin{abstract}

We demonstrate a novel technique for the measurement of the charge
carried by a colloidal particle. The technique uses the phenomenon
of the resonance of a particle held in an optical tweezers trap and
driven by a sinusoidal electric field. The trapped particle forms a
strongly damped harmonic oscillator whose fluctuations are a
function of $\gamma$, the ratio of the root-mean square average of
the electric and thermal forces on the particle. At low applied
fields, where $\gamma \ll 1$, the particle is confined to the
optical axis while at high fields ($\gamma \gg 1$) the probability
distribution of the particle is double-peaked. The
periodically-modulated thermal fluctuations are measured with
nanometer sensitivity using an interferometric position detector.
Charges, as low as a few elementary charges, can be measured with an
uncertainty of about 0.25 $e$. This is significantly better than
previous techniques and opens up new possibilities for the study of
nonpolar suspensions.

\end{abstract}

\maketitle

\section{Introduction}

Although charge is often thought to play no role in nonpolar media
there is ample \cite{3192,3305,3771}, if sometimes contradictory
\cite{Morrison-868}, evidence of surface charging in nonpolar
colloidal suspensions. Much of the reason for this uncertainty is
the extremely low level of charge in nonpolar environments. While a
particle of radius $a \sim 500$ nm might carry a surface charge
$eZ_{\rm{eff}}$ of the order of 10$^{3}$ electrons in an aqueous
dispersion ($\epsilon_{r} \sim 80$), the charge in a nonpolar
solvent will be some two or three orders of magnitude smaller.
Although small in absolute terms, this charge still produces
surprisingly strong effects in low dielectric environments. For
instance, using Coulomb's law the contact value of the interaction
potential (in units of $k_{B}T$) between two colloidal spheres with
radius $a$ and charge $eZ_{\rm{eff}}$ in a solvent of dielectric
ratio $\epsilon_{r}$ is simply,
\begin{equation}\label{eqn-contact_pot}
    \frac{U_{0}}{k_{B}T}  =  \frac{1}{k_{B}T} \cdot \frac{Z_{\rm{eff}}^{2}e^{2}}{8\pi \epsilon_{0}
    \epsilon_{r} a}     =  \left ( \frac{\lambda_{B}}{2a} \right )
    Z_{\rm{eff}}^{2},
\end{equation}
where we have introduced the Bjerrum length $\lambda_{B} = e^{2} / 4
\pi \epsilon_{0} \epsilon_{r} k_{B}T$ as a characteristic length
scale of electrostatic interactions. For dodecane, $\lambda_{B}$ is
28.0 nm so the repulsive interactions between 500 nm particles,
carrying a typical charge of 10 electrons, is $\approx 3 \; k_{B}T$
-- big enough to have a dramatic effect on the structure of a
colloidal suspension.

The most direct route to monitor the extent of particle charging is
to measure the uniform drift velocity $u_{D}$  in the presence of a
weak electric field $E$. For a spherical particle, $u_{D} = \mu E$,
where $\mu$ is the electrophoretic mobility. For an isolated
colloidal sphere, in the low-salt regime, moving in a solvent of
viscosity $\eta$, Stokes' law implies $\mu = Z_{\rm{eff}}e /(6 \pi
\eta a)$ since the ionic atmosphere around the colloid is very
diffuse. The low charges found in nonpolar media therefore translate
to electrophoretic mobilities some two or three orders of magnitude
or so less than in water, at the same electric field. Rapid and
reproducible measurements of such low mobilities presents a major
challenge to conventional electrokinetic techniques \cite{3192}.

Motivated by the growing acknowledgement of the importance of charge
in nonpolar suspensions \cite{3067} and its technological
significance \cite{3444}, several novel methods have been developed
to determine particle mobility in nonpolar suspensions. Sch\"{a}tzel
and coworkers \cite{548} pioneered an innovative amplitude-weighted
phase analysis signal processing scheme for the analysis of
classical laser Doppler electrophoresis data. The technique, phase
analysis light scattering (PALS), is significantly more sensitive
than conventional laser Doppler electrophoresis. In practice
however, collective particle motion induced by thermal convection or
sedimentation restricts the achievable sensitivity of commercial
PALS instruments  to $\sim$ 10$^{-10}$ m$^{2}$ s$^{-1}$ V$^{-1}$,
which corresponds to colloid charges of $\sim 10e$ on a particle of
radius $a = 500$ nm. Video microscopy has been used by several
authors. Perez and Lemaire \cite{3527} used a vibrating
piezoelectric stage to null the oscillatory microscopic motion
produced by an applied alternating field, while Strubbe et al.
\cite{4438} have measured the diffusion constant and electrophoretic
mobility of individual particles in a nonpolar suspension. Despite
these developments there is still no technique which can provide
fast and accurate measurement of the extremely small levels of
particle charge found in  nonpolar suspensions. As a result the
microscopic charging mechanisms in non-aqueous media has remained
largely problematic \cite{Morrison-868}. Improvements in
quantitative characterization are an essential first step to
developing a more detailed understanding of the role of charge in
nonpolar suspensions.

In this paper, we present a simple technique that allows a direct
measurement of the effective charge of colloidal particles in
nonpolar solvents. Our approach is very different from standard
electrophoretic light scattering methods and uses the phenomenon of
the driven resonance of a strongly-damped harmonic oscillator. We
use the radiation pressure from a laser beam focused to a near
diffraction-limit spot (an `optical tweezer') to trap single
colloidal particles in a dilute suspension. The focused beam
produces a harmonic potential for the trapped particle whose
strength is proportional to the laser power. The hindered diffusion
of the trapped colloid is modified by an oscillatory electric field.
The modulated thermal fluctuations are measured with nanometer
sensitivity and the charge extracted from the spectral density of
the fluctuations at the field frequency.

Our approach is similar in principle to the classical Millikan oil
drop experiments \cite{4527} but uses light rather than gravity. In
his famous experiment to prove the quantization of charge, Millikan
measured the speed of a charged oil drop under gravity and its speed
falling in a electric field $E$. He switched on electric forces
$F_{E} = ZeE$ of between 300--1300 fN (1 fN = 10$^{-15}$N) on oil
drops with radii between 2 and 6 $\mu$m (each drop containing $Z$
elementary charges, $e$) to counterbalance the gravitational force
$mg$ acting on the drop. Our experiments rely, instead, on using the
random thermal forces leading to Brownian motion, to oppose an
applied electric force. The balance of forces determines the
diffusive motion of the particle, rather than the velocity of
free-fall as in Mullikan's experiment. Below we demonstrate that, by
using a weak optical trap and an interferometric position detection
system, we can measure accurately a mean particle charge of -2.9\;e
which, for the typical field strengths used, corresponds to an
electric force of about 40 fN. We estimate that the sensitivity of
our technique is about an order of magnitude smaller than this
limit. We believe our method is a significant improvement on
existing techniques. It provides rapid, accurate and highly
sensitive information on the status of particle charging in nonpolar
suspensions. Related but less sensitive experiments have been
described on highly-charged aqueous colloids by Garbow et al.
\cite{4228} where the authors used a weakly-focused laser beam. This
had the effect of pushing particles along the optical axis, rather
than stably confining them, as in our approach.

This paper is organized as follows: After a description of the
apparatus and measurement procedures, we give a short introduction
to the theory of a driven overdamped harmonic oscillator before
results obtained on a variety of different nonpolar suspensions are
presented and discussed in the final section.

\section{Materials and Methods} \label{sec-experiments}

\subsection{Materials.}

We use a model non-aqueous colloidal system of sterically-stabilized
poly(methyl methacrylate) spheres of radius $a = 610$ $\pm$ 30 nm
suspended in dodecane. The particles were synthesized by dispersion
polymerization, following the procedures detailed by Antl et al.
\cite{Antl-63}. Electron microscopy revealed the particles were
highly uniform in size with a radius polydispersity (root mean
square variation / mean radius) of $\sigma = 0.046$ $\pm$ 0.01.
Dodecane was dried with activated molecular sieves (Acros, size 4A)
and stored under dry nitrogen.  Sodium bis(2-ethylhexyl)
sulfosuccinate (NaAOT, Fluka BioChemika Ultra 99\%) was purified by
dissolution in methanol and tumbled with activated charcoal.
Zirconyl 2-ethyl hexanoate [Zr(Oct)$_{2}$] was purchased from Alfa
Aesar (Heysham, UK) and used as received.

\subsection{Optical Tweezers.}

Figure~\ref{fig:tweezer-schematic} shows a schematic of our optical
tweezer system. A fibre-coupled laser beam with wavelength $\lambda
= 1064$ nm (Ytterbium fiber laser, IPG Photonics, Germany) was
focussed to a diffraction-limited beam waist by a microscope
objective (Plan-Neofluar, $\times$100, N.A. 1.3, Zeiss) mounted in
an inverted microscope (Axiovert S100, Zeiss).  The laser intensity
($\sim$ 10 mW) was varied using a $\lambda /2$ waveplate and a
polarizing beam-splitter cube placed in the beam path. The
fluctuations in position of the trapped particle, with respect to
the center of the trap, were measured with a quadrant photodetector
(model QD50-4X, Centronics, UK) using back-plane optical
interferometry \cite{1614}. The particle positions were acquired by
a Labview programme (National Instruments, Austin, Texas, USA) and
digitized using a high-speed data acquisition card (National
Instruments model PCI-MI0-16E-1) at 10 kHz. For the current
experiments, the number of particle positions collected in each time
trace was set at 2$^{18}$, so that the duration of each measurement
was $\approx 26$ s.  Positions measured in voltages were converted
into displacements in nanometers by recording the time-dependent
mean-square displacement $\left < \Delta x^{2}(\tau) \right >$  of
five particles from the same batch of particles, immediately before
each set of electrophoresis measurements.  The laboratory
temperature was stabilized at 22 $\pm$ 1 \degc. At this temperature
the viscosity of dodecane is 1.378 x 10$^{-3}$ Pa s and the
dielectric constant is 2.002.

\subsection{Single Particle Optical Microelectrophoresis (SPOM).}

The microelectrophoresis cell was constructed from two platinum foil
electrodes, 127 $\mu$m high and with a width of 2 mm, mounted in a
cylindrical glass sample chamber, 1.13 mm high with a volume of
$\sim 90 \; \mu$l. The cell was constructed in three stages. First,
two platinum foil electrodes were carefully positioned parallel to
each other onto a 5.0 cm x 2.5 cm rectangular glass coverslip (170
$\mu$m thick), covered with a thick layer of a UV-activated glue
(Loctite 350), using a low-magnification stereo microscope. The
electrodes were securely fixed in place by irradiation with a 100 W
UV-lamp. Then a 1 mm thick glass slide with an 1 cm diameter inset
circular hole was placed on top of the electrodes and sealed to the
electrodes with UV glue. Finally, immediately prior to use, the top
of the chamber was closed by a 22 mm diameter 170 $\mu$m-thick
circular coverslip. The use of optical quality cover-slips on both
the bottom and top of the sample chamber minimized spherical
abberation.

The distance between the two electrodes was measured, after assembly
of the cell, using a calibrated microscope graticule as 128 $\mu$m
$\pm$ 2$\mu$m. The small gap between the electrodes makes it
possible to generate electric fields on the order of tens of
kilovolts per meter with just a few volts applied to the cell, which
minimizes thermal instabilities due to Joule heating and avoids the
electrohydrodynamic fluid instabilities which occur at higher
voltages \cite{Novotny}. The electric field was estimated from the
expression $E = \lambda V/d$, where $V$ is the applied voltage, $d$
is the plate separation and, $\lambda$ is a factor correcting for
the finite size of the electrodes. Analytical calculations
\cite{Morse}, for the field on the midplane equidistant between two
semi-infinite plates, give $\lambda = 0.969$ for the rectangular
electrodes used. Accumulation of ions near the electrode and the
generation of a non-uniform space charge was suppressed by using an
ac voltage. Electrosmosis was minimized by positioning the
electrodes 60 $\mu$m above the bottom glass wall, so that the
tangential electric field at the wall was reduced. Moving the
position of the trapped particle away from the electrode center by
10 $\mu$m, in any direction, changed the recorded charge by less
than 2.5\%, confirming the absence of electrosmotic fluid flows.

The sample chamber was filled with a dilute suspension of colloidal
particles with a volume fraction of $\sim 10^{-5}$. The cell was
carefully centered so that the laser passed through a plane
equidistant between the two electrodes and the focus was midway
between the top and bottom surfaces. The particle was held at least
65 $\mu$m from the nearest surface, to ensure Stokes' law was
applicable. The charge on the particle was analysed using a
purpose-written software package written in IDL (Research Systems,
Boulder, Colorado). The sign was determined by reducing the
frequency of the applied electric field to 0.5 Hz, blocking the
laser beam so that the particle was momentarily released from the
optical trap, and following the oscillatory motion of the free
particle on a CCD camera.

The accuracy of SPOM was checked by measuring the
mobility of a 850 nm (radius) PMMA particle in dodecane (with 100 mM
NaAOT) using a commercial PALS instrument (Brookhaven ZetaPlus),
$\mu_{\rm{PALS}}$, and the SPOM technique, $\mu_{\rm{SPOM}}$. The
good agreement between $\mu_{\rm{SPOM}}$ (($-5.4 \pm$ 0.5) x
10$^{-10}$ m$^{2}$ s$^{-1}$ V$^{-1}$) and $\mu_{\rm{PALS}}$ (($-6.0
\pm$ 0.5) x 10$^{-10}$ m$^{2}$ s$^{-1}$ V$^{-1}$) confirms that the
optical microelectrophoresis technique yields accurate particle
mobility data.

\section{Theory}

A colloidal particle illuminated by a laser beam encounters a
gradient force that is oriented in the direction of the intensity
gradient and a scattering force that is oriented in the direction of
the incident light. In a tightly-focused beam the gradient forces
dominate and lead to a strong restoring force, which causes the
particle to be confined near the focus of the beam. To a first
approximation, the particle can be modeled as a mass $m$ in a
three-dimensional harmonic potential. The potential is characterized
by two spring constants, one in the axial and one in the radial
direction. Here, for simplicity, we analyze only the one-dimensional
motion along the radial coordinate $x$, where the corresponding
force constant is $k_{H}$.

Placed in an external electric field, the total force on a
\textit{charged} particle is a sum of three contributions: a
harmonic force, $-k_{H} x(t)$, arising from the optical trap; a
random force $\mathcal{R}(t)$ that represents thermal forces at the
temperature $T$; and an external periodic force $F_{p}(t) = A
\sin(\omega_{p} t + \phi)$ which is characterized by the frequency
$\omega_{p}$, amplitude $A$ and initial phase $\phi$ of the external
field. Defining the effective charge on the particle as
$Z_{\rm{eff}}$ (in units of the fundamental charge $e$) then the
amplitude of the oscillatory force is
\begin{equation}\label{eqn-amplitude}
    A= Z_{\rm{eff}} e E,
\end{equation}
where $E$ is the maximum electric field strength at the particle
center.  In a single particle optical microelectrophoresis
experiment, the quantity ultimately measured is the time-dependent
particle autocorrelation function $C(\tau)$, or its
Fourier-transform, the spectral density $I(\Omega)$. Below we derive
expressions for both quantities, in terms of the effective charge
$Z_{\rm{eff}}$.

\subsection{A periodically-driven Brownian Oscillator} \label{sec-Brownian oscillator}

The fluctuations of a periodically-driven Brownian oscillator are
described, in the conventional Ornstein-Uhlenbeck theory
\cite{Uhlenbeck-652}, by the Langevin equation
\begin{equation}\label{eqn-langevin}
    m \ddot{x}(t) + \xi \dot{x}(t) + k_{H} x(t) = A \sin(\omega_{p} t + \phi) +
    \mathcal{R}(t).
\end{equation}
Here $x(t)$ is the particle trajectory, and $\xi = 6 \pi \eta a$ is
the friction coefficient of a sphere of radius $a$ moving in a
medium of viscosity $\eta$. The random thermal forces are modeled by
a Gaussian process $\mathcal{R}(t)$ with the moments,
\begin{equation}\label{eqn-noise}
    \left < \mathcal{R}(t) \right > = 0, \hspace{0.25in} \mbox{and}
    \hspace{0.25in}
    \left < \mathcal{R}(t) \mathcal{R}(t^{\prime})\right
    > = 2 \xi k_{B}T \delta(t-t^{\prime}).
\end{equation}
This definition ensures that the Brownian particle is in thermal
equilibrium at a temperature $T$, in the absence of any driving
field i.e. for $A=0$.

A colloidal particle, once moving, loses momentum rapidly as a
result of viscous losses. The characteristic time for this decay is
$t_{B} = m/\xi$, which is about 10$^{3}$ times smaller than the time
resolution of our experiments, when digitizing at a 10kHz sampling
rate. Consequently the inertial terms may safely be dropped so that
Eq.~\ref{eqn-langevin} now reads
\begin{equation}\label{eqn-overdamped}
        \xi \dot{x}(t) + k_{H} x(t) = A \sin(\omega_{p} t + \phi) +
    \mathcal{R}(t).
\end{equation}
 The equation  is linear so that the general solution  may be
written \cite{4466},
\begin{equation}\label{eqn-superposition}
    x(t) = x_{n}(t) + x_{p}(t),
\end{equation}
where $x_{n}$ is the solution in the presence of  random thermal
forces \textit{only} (i.e. $A=0$) and $x_{p}$ is the solution when
only periodic forces act (i.e. $T=0$). We now consider each solution
in turn.

The driven oscillator is deterministic in the absence of any random
thermal noise. The solution is well known \cite{4466},
\begin{equation}\label{eqn-xp}
    x_{p}(t)  =  \frac{A}{k_{H} \left [ 1 + (\omega_{p} / \omega_{c})^{2}
    \right ]^{1/2}} \sin (\omega_{p} t + \phi - \Delta ).
\end{equation}
The particle motion lags the driving field by a constant phase
factor,
\begin{equation}\label{eqn-Delta}
        \Delta  =  \tan^{-1} \frac{\omega_{p}}{\omega_{c}},
\end{equation}
which varies with the relative sizes of the driving and the
\textit{corner frequency},
\begin{equation}\label{eqn-corner-freq}
    \omega_{c} = k_{H} / \xi.
\end{equation}
The frequency ratio $\omega_{p} / \omega_{c}$  also controls the
amplitude of the particle motion. For low frequencies ($\omega_{p}
\ll \omega_{c}$), the amplitude is essentially frequency
independent. While, at high frequencies ($\omega_{p} \gg
\omega_{c}$) the response decreases rapidly with increasing
$\omega_{p}$. In between, Eq.~\ref{eqn-xp} reveals a monotonic
oscillation spectrum, without the dynamic resonance at $\omega_{p} =
\omega_{c}$, characteristic of underdamped systems. The relative
phase of the particle motion depends on the initial phase $\phi$ of
the applied force. In our experiments $\phi$ is a fluctuating
stochastic variable with a coherence time $\tau_{p}$, which depends
on the stability of the voltage generator. The phase is essentially
randomized on times $t \gg \tau_{p}$ so time and phase averages are
equivalent. Multiplying the trajectories at $t$ and $t^{\prime}$ and
averaging over  $\phi$, assuming the initial value of $\phi$ is
uniformly distributed over the interval $[0,2 \pi]$, yields  the
correlation function,
\begin{equation} \label{eqn-correlation-periodic}
 \left < x_{p}(t) x_{p}(t^{\prime}) \right >_{\phi}  = \frac{A^{2}}{2k_{H}^{2} \left [ 1 + (\omega_{p} / \omega_{c})^{2}
    \right ]} \cos \omega_{p} (t- t^{\prime}),
\end{equation}
where the subscript denotes the phase-averaging.  Since the motion
is totally deterministic, $\left < x_{p}(t) x_{p}(t^{\prime}) \right
>_{\phi}$ oscillates continuously between positive and negative values.

The dynamics of the trapped particle is a linear superposition of
periodic and random motion. The correlation function of a purely
Brownian oscillator is given by Doi and Edwards \cite{Doi-646} as,
\begin{equation}\label{eqn-correlation-noise}
\left < x_{n}(\tau) x_{n}(0)\right
>_{\mathcal{R}} = \frac{k_{B}T}{k_{H}} \exp ( -\omega_{c} \tau).
\end{equation}
The mean-square displacement is accordingly $\left < x_{n}^{2}
\right
>_{\mathcal{R}} = k_{B}T / k_{H}$ and, since $F=-k_{H} x$, the mean-square thermal
forces on the trapped particle are $\left < F_{n}^{2} \right
>_{\mathcal{R}} = k_{B}T k_{H}$.

\subsection{Correlation function and spectral density}
The correlation function $c(\tau)$ of the periodically-driven
Brownian oscillator is defined by
\begin{equation}\label{eqn-total-correlation}
    c(\tau = t^{\prime}-t) = \left < \left < x(t) x(t^{\prime}) \right
    >_{\mathcal{R}} \right >_{\phi},
\end{equation}
where the averages are taken over both noise and phase. Since the
thermal and periodic forces are uncorrelated, $c(\tau)$ is simply a
sum of the correlation functions for periodic and purely thermal
motion,
\begin{equation}\label{eqn-total-corr}
    c(\tau) = \frac{k_{B}T}{k_{H}} \exp ( -\omega_{c} \tau) + \frac{A^{2}}{2k_{H}^{2} \left [ 1 + (\omega_{p} / \omega_{c})^{2}
    \right ]} \cos \omega_{p} \tau.
\end{equation}
It is more convenient to work with the normalized function $C(\tau)
= c(\tau)/c(0)$, which may be written as
\begin{equation}\label{eqn-norm-correlate}
    C(\tau) = \frac{1}{1+ \gamma^{2}}\exp ( -\omega_{c} \tau) + \frac{\gamma^{2}}{1+
    \gamma^{2}} \cos \omega_{p} \tau,
\end{equation}
where we have introduced $\gamma^{2}$, the scaled {\textit {ratio of
the mean-square periodic and Brownian forces},
\begin{equation}\label{eqn-gammasq}
    \gamma^{2} = \frac{\left < F_{p}^{2} \right > / \left < F_{n}^{2} \right
    > }{1 +
    (\omega_{p}/\omega_{c})^{2} }.
\end{equation}
Here $\left < F_{p}^{2} \right > = Z_{\rm{eff}}^{2} e^{2} E^{2} / 2$
and $\left < F_{n}^{2} \right> = k_{B}T k_{H}$. The force ratio
$\gamma$ appears naturally in the theory of a driven Brownian
oscillator and, as we show below, is also the quantity most readily
extracted from experiment. Knowing $\gamma$, the charge on the
particle follows immediately as,
\begin{equation}\label{eqn:charge-from-gamma}
    e |  Z_{\rm{eff}} | = \frac{\gamma \xi }{E} \sqrt{\frac{2k_{B}T}{k_{H}}
    (\omega_{p}^{2}+\omega_{c}^{2})}.
\end{equation}
This expression simplifies at low frequencies ($\omega_{p} \ll
\omega_{c}$) to,
\begin{equation}\label{eqn-charge-low-freq}
e | Z_{\rm{eff}} | =  \frac{\gamma }{E} \sqrt{2 k_{B}T k_{H}}
\end{equation}
which does not depend on the radius of the trapped particle.

Figure~\ref{fig:curvature} shows the force ratio $\gamma$ calculated
for typical experimental parameters. In the weak-field limit, where
$\gamma \ll 1$, the motion of the trapped particle is dominated by
random thermal forces. By contrast, in the strong-field limit where
$\gamma \gg 1$, the thermal forces are only a relatively small
perturbation and the electrophoretic forces dominate. The cross-over
between these two regimes occurs at a particle charge of
$Z_{\rm{eff}} \sim 20$, for typical field strengths. At this point
the electric forces on the particle are a few hundred femtonewtons.

The force ratio $\gamma$ may be determined experimentally from the
correlation function $C(\tau)$ or, equivalently, from the spectral
density,
\begin{equation}\label{eqn-Wiener-Khintchin}
    I(\Omega) = \frac{1}{2 \pi} \int_{-\infty}^{\infty} c(\tau)
    \exp (-i \Omega \tau) d\tau.
\end{equation}
As we discuss below there are advantages to both approaches.
Inserting Eq.~\ref{eqn-total-corr} into \ref{eqn-Wiener-Khintchin}
reveals that the spectral density is a sum of two
independent contributions,
\begin{equation}\label{eqn-power-spectrum}
  I(\Omega) = \frac{k_{B}T}{\pi \xi} \frac{1}{\Omega^{2} +
  \omega_{c}^{2}} + \frac{k_{B}T \gamma^{2}}{2k_{H}} [ \delta(\Omega
  - \omega_{p}) + \delta(\Omega + \omega_{p})],
\end{equation}
a Lorentzian spectrum typical of Brownian motion in a harmonic
potential, together with additional $\delta$-spikes at the positive
and negative field frequencies. Integrating the spikes over the
frequency axis gives an expression for the mean-squared periodic
displacement of the particle $P_{sig}$,
\begin{equation}\label{eqn-Psignal}
    P_{sig} = \int_{-\infty}^{\infty} I_{p}(\Omega)d\Omega =
    \frac{k_{B}T}{k_{H}}\gamma^{2}.
\end{equation}
from which $\gamma$ can be found.

\subsection{Asymptotic probability distribution function}
\label{sec:pdf-theory}

We now turn to the probability distribution of the confined
particle. Introducing the dimensionless position coordinate $\bar{x}
= x /(k_{B}T/k_{H})^{1/2}$, we define the probability of locating
the Brownian particle in the region between $\bar{x}$ and
$\bar{x}+d\bar{x}$ as $p(\bar{x})d\bar{x}$, where $p(\bar{x})$ is
the probability distribution function.

Without a periodic force, the Brownian particle is at thermal
equilibrium and $\bar{x}(t)$ follows the Boltzmann's distribution,
\begin{equation}\label{eqn-pdf}
    p_{n}(\bar{x}) = \frac{1}{\sqrt{2 \pi}} \exp (- \bar{x}^{2} /2
    ).
\end{equation}
Applying a time-dependent force perturbs the system and shifts the
average distribution away from the equilibrium result. When all
transients have died out, the probability approaches a new
asymptotic distribution which is a periodic function of the time
$t$. As discussed earlier, the only experimentally accessible
quantities are the phase-averaged functions. Analytical expressions
for the phase-averaged distributions have been proposed by Jung and
H\"{a}nggi \cite{4168}, for arbitrary friction coefficients. In the
overdamped limit, their result reduces to the infinite series,
\begin{widetext}
\begin{equation}\label{eqn-general-pdf}
    p(\bar{x}) = \frac{1}{\sqrt{2 \pi}} \exp \left ( -\frac{\bar{x}^{2} +
    \gamma^{2}}{2} \right ) \left \{ I_{0}\left (\frac{\gamma^{2}}{2} \right)
    I_{0} \left( \sqrt{2} \gamma \bar{x} \right) + 2 \sum_{k=1}^{\infty} (-1)^{k}
    I_{k} \left (\frac{\gamma^{2}}{2} \right) I_{2k} \left( \sqrt{2} \gamma \bar{x} \right)
    \right \}
\end{equation}
\end{widetext}
where  $I_{k}(x)$ is a modified Bessel function \cite{4468}.
Figure~\ref{fig:theoretical pdfs} shows the effect of a finite
external field on the probability distribution. With increasing
amplitude, the probability $p(\bar{x})$ first broadens until, at
$\gamma = \gamma^{\ast}$, the distribution becomes bimodal and
develops two peaks around $\bar{x} \sim \pm\; \gamma$.  Further
increases in the external field lead the two peaks to split further
apart, leaving a relatively flat distribution around the origin, as
the particle becomes increasingly localized around the two
extremities.

The changes in the probability distribution with external field can
be clearly seen from the curvature of $ p(\bar{x})$ near to the
origin. Expanding Eq.~\ref{eqn-general-pdf} around $\bar{x}=0$
gives, to terms in $\bar{x}^{2}$,
\begin{equation}\label{eqn-power-expansion}
 p(\bar{x}) = \frac{1}{\sqrt{2 \pi}} \exp \left ( -\frac{
    \gamma^{2}}{2} \right ) I_{0}\left (\frac{\gamma^{2}}{2} \right)
    (1-\kappa \bar{x}^{2})
\end{equation}
where the curvature is
\begin{equation}\label{eqn-kappa}
    \kappa = \frac{1}{2} \left [1- \gamma^{2}+\gamma^{2} \frac{I_{1}(\gamma^{2}/2)}{I_{0} ( \gamma^{2} /2 )}
    \right ].
\end{equation}
$\kappa$ is positive for a single monomodal distribution and
negative for a bimodal distribution. The inset plot in
Fig.~\ref{fig:theoretical pdfs} displays the variation of $\kappa$
with  external field. The curvature first decreases rapidly with
increasing $\gamma$, reflecting the broadening of $p(\bar{x})$
evident at low modulation strengths, before changing sign for
$\gamma
> \gamma^{\ast} = 1.257$ as the distribution becomes bimodal.
The small negative values of the curvature found at high fields
correspond to the extremely flat shape of the distribution seen
close to the origin at high $\gamma$. The curvature is a function
solely of $\gamma^{2}$ so that the distribution sharpens
continuously with increasing frequency and does not show the
resonant bimodality evident in the underdamped limit \cite{4168}.

The width of the particle distribution is  $\left < x^{2} \right
> $, which from Eq.~\ref{eqn-total-corr}, equals
\begin{equation}\label{eqn-msd}
\left < x^{2} \right
    > =
    \frac{k_{B}T}{k_{H}}(1+\gamma^{2})
\end{equation}
where $\left< \cdots \right >$ denotes an average over both
$\mathcal{R}$ and $\phi$. In the weak-field limit where $\gamma \ll
1$, the width of the distribution approaches the thermal
limit, $k_{B}T / k_{H}$. Broadening of the probability distribution
is only significant in the strong-field limit ($\gamma \gg 1$) where
the width varies quadratically with the strength of the applied
field.

\section{Results and Discussion}

\subsection{Field and Frequency Dependence of Effective Charge}

To validate our technique, experiments were first carried out on a
number of weakly-charged particles, as a function of the applied
field strength and its frequency. The particles were taken from a
dilute dispersion of 610 nm (radius) poly(methyl methacrylate)
particles in a 0.035 wt\% solution of a PHSA-g-PMMA copolymer in
dodecane. Figure~\ref{fig:exp-trajectory} shows a short portion of
the $x$- and $y$-position of a trapped particle when a sinusoidal
field is applied. The field was applied along the $x$-axis so
periodic oscillations are expected in the $x$-signal and should be
absent from the $y$-signal. Looking at the trajectories plotted in
Fig.~\ref{fig:exp-trajectory}, it is very difficult to see any
significant differences between the $x$- and $y$-signals. The small
electrophoretic oscillations expected are `buried' beneath much
larger Brownian fluctuations and $\gamma \ll 1$. To extract the
small periodic excursions from the predominantly random Brownian
motion we determine the spectral distribution of the fluctuations.

Figure~\ref{fig:autocorrelations} shows the normalized
autocorrelation functions, $C(\tau)$, measured on the same particle,
as the applied field was varied. The oscillatory component is
clearly visible, particularly at long times. Data collection is fast
with each of the functions shown representing just 26 s worth of
data. The alternating electric field had a frequency of 67 Hz and
voltages of up to 10 V were applied to the electrodes. The laser
trap had a stiffness of $k_{H}= 4.7 \pm 0.1$ fN nm$^{-1}$ and a
corner frequency of 47 $\pm$ 1 Hz.  The agreement between the
experimental data and the least-squares fits is excellent.
Fig.~\ref{fig:autocorrelations}(a) reveals that at short-times the
correlation functions all essentially converge, even though their
long time dependence is clearly very different. This is because on
timescales shorter then the period of the applied field, the motion
of the trapped particle is dominated by Brownian diffusion. Thermal
fluctuations are independent of the field and, since they are
uncorrelated, result in the exponential relaxation apparent at short
times in $C(\tau)$ (Eq.~\ref{eqn-correlation-noise}). The
oscillations evident at long times
(Fig.~\ref{fig:autocorrelations}(b)) reflect the periodic motion
produced by the applied field and accordingly grow as the applied
field is increased.

Fitting the measured correlation functions to the theoretical
expression for $C(\tau)$ (Eq.~\ref{eqn-norm-correlate}) yields
$\gamma$, the ratio of the rms electrophoretic and Brownian forces.
The ratio is plotted in Figure~\ref{fig:field-dependence} as a
function of the applied field $E$. As expected, the experimental
values of $\gamma$ are linearly proportional to $E$, confirming that
the charge on the particle is constant.  The gradient of the plot
yields an effective charge of $| Z_{\rm{eff}} | = 14.5 $ $\pm$ 0.2
$e$, equivalent to an electrophoretic mobility of $\mu$ = 1.46 $\pm$
0.02 x 10$^{-10}$ m$^{2}$ s$^{-1}$ V$^{-1}$. The charge on this
particle is sufficiently low that, even at the maximum field
strengths achievable, the system remains in the low-field limit
($\gamma < 1$) where  Brownian forces dominate.

For ultra-low charges where $\gamma \ll 1$ it becomes increasingly
difficult to determine the oscillatory element in the
autocorrelation function, because of fluctuations in the baseline.
This can be seen in the data at 13.3 kV m$^{-1}$ in
Figure~\ref{fig:autocorrelations}(b)
 where the oscillations  are indistinct. To
resolve these low particle charges we perform a fast Fourier
transform of the data and calculate the spectral density.
Figure~\ref{fig:exp-power} shows the field dependence of the
resulting power spectra. The measured spectra are particularly
simple, being a superposition of a Lorentzian spectrum (shown
dashed) reflecting diffusive motion and a single $\delta$-peak at
the fundamental electrode drive frequency $\omega_{p}$, whose height
increases with the applied field. The data contains no higher
harmonics of $\omega_{p}$ confirming the linear response of particle
and field. To extract the power in the periodic signal, the data
around the $\delta$-spike was masked and the diffusive spectrum
fitted to the first term of Eq.~\ref{eqn-power-spectrum}, by
adjusting the unknown corner frequency $\omega_{c}$. Subtracting the
fitted Lorentzian from the measured power spectra yielded the signal
spectrum. The mean-square electrophoretic displacement $P_{sig}$ was
obtained by integrating the signal spectrum around the peak at
$\omega_{p}$. The resulting values for $P_{sig}$ are plotted as a
function of the square of the applied field, in the inset of
Figure~\ref{fig:exp-power}. As expected from Eq.~\ref{eqn-Psignal},
the mean-square displacement $P_{sig}$ varies linearly with $E^{2}$.
The high positional sensitivity achievable is highlighted by the
arrowed $\delta$-peak in Fig.~\ref{fig:exp-power} which, while it
remains clearly visible, corresponds to a mean-square displacement
of only 1.4 nm$^{2}$.

To explore how the frequency of the applied field influenced the
measurements we collected autocorrelations at a number of different
field frequencies, taking care to ensure that the same particle was
used in each case. The corner frequency of the trapped particle was
$\omega_{c} = 291$ rad s$^{-1}$ and data was recorded at frequencies
between 25 rad s$^{-1}$ and 1930 rad s$^{-1}$, approximately an
order of magnitude below and above the corner frequency of the
optical trap. The amplitude of the applied field was fixed at 75.8
kV m$^{-1}$. The autocorrelations were analyzed as described above
to determine $\gamma^{2}$ at each frequency. The results are
summarized in Figure~\ref{fig:exp-freq}. As expected, from the
discussion in \ref{sec-Brownian oscillator}, the response of the
trapped particle to the applied field decreases
 with increasing frequency. This drop in sensitivity is particularly sharp when the driving
frequency exceeds the corner frequency of the optical trap. Plotting
$\gamma^{2}$ as a function of the frequency ratio $1 +
(\omega_{p}/\omega_{c})^{2}$ confirms that $\gamma^{2}$ is inversely
dependent on $1 + (\omega_{p}/\omega_{c})^{2}$, in agreement with
the predictions of Eq.~\ref{eqn-gammasq}. The highly linear
dependency evident in Fig.~\ref{fig:exp-freq} confirms the validity
of the driven harmonic model and indicates that the charge on the
particle is frequency independent. Analysis gives the surface charge
on the particle as $|Z_{\rm{eff}}| = 15.8$ $\pm$ 0.2.

\subsection{Field Dependence of Probability Distribution Function}

So far, we have found excellent agreement between our measurements
and the theoretical predictions for a
periodically-driven oscillator. The comparison has however
been limited to weakly-charged particles where $\gamma < 1$. To
extend our understanding, we now look at more highly-charged
particles where strong electrophoretic forces dominate the weaker
Brownian forces. Hsu et al. has shown \cite{3771} that non-polar
suspensions may be charged by adding the surfactant, sodium
bis(2-ethylhexyl) sulfosuccinate (Na-AOT), which forms reverse
micelles in a non-polar solvent. Adding Na-AOT to our suspensions at
concentrations of 100 mM increases the average charge (in electrons)
on the PMMA particles by a factor of  four to $Z_{\rm{eff}} \approx
- 50$. Using these more highly-charged particles we are able to
investigate the high-field limit where $\gamma > 1$.

Figure~\ref{fig:exp-pdf} shows the field dependence of the particle
probability. The normalized distribution function is plotted as a
function of the dimensionless position $\bar{x} = x
/(k_{B}T/k_{H})^{1/2}$, where the $x$-location of the sphere has
been scaled by the size of the corresponding \textit{rms} Brownian
fluctuations in the optical trap. To correct for small asymmetries
in the trap caused by optical misalignment, the measured probability
was averaged together at each positive and negative displacement.
The resulting symmetrized distributions, $p_{s}(\bar{x})=
[p(\bar{x}) + p(-\bar{x})] / 2$, are plotted in
Figure~\ref{fig:exp-pdf}. At zero field the probability distribution
is a simple Gaussian function, as expected from Eq.~\ref{eqn-pdf}.
Increasing the field to $E \approx 23$ kV m$^{-1}$ causes the
distribution to broaden, although it remains singly-peaked. At
fields of 46 kV m$^{-1}$ and above there is a dramatic change in
$p_{s}(\bar{x})$. The central peak in the distribution first splits
into two broad maxima, before then moving further apart in $\bar{x}$
as the field is increased still further. Using the
experimentally-determined charge on the particle ($Z_{\rm{eff}} =
-48.3 \pm 0.3$) and the measured optical trap stiffness we
calculate the force ratio $\gamma$ from the applied
electric field $E$. In this way we have confirmed that the
probability distribution first becomes bimodal at a force ratio in
the range $1.01 < \gamma < 1.34$. This observation is in good
agreement with the theoretical predictions of
Section~\ref{sec:pdf-theory} where a bimodal distribution is
predicted for $\gamma > \gamma^{\ast} = 1.257$.

To  account for the dramatic changes evident in the measured
distributions we have calculated $p(\bar{x})$ for a
periodically-driven oscillator. The exact shape of this distribution
depends solely upon the force ratio $\gamma$, as noted in
Section~\ref{sec:pdf-theory}.  Fixing $\gamma$ at the experimental
value, we have evaluated numerically the series of Bessel functions
given in Eq.~\ref{eqn-power-expansion}. The infinite series was
truncated when convergence to four significant figures was achieved.
The results of these calculations are plotted as the solid lines in
Figure~\ref{fig:exp-pdf}. The agreement is remarkably good and is
particularly encouraging given that there are no adjustable
parameters in the comparison.

\subsection{Sensitivity Limit}

The smallest charge that can be detected by our technique is fixed
by the point where the peak in the spectral density at $\omega_{p}$
(the arrowed peak in Fig.~\ref{fig:exp-power}) becomes
indistinguishable from the Brownian background. At this point the
spectral power in a band of width $\Delta \omega$ centred on
$\omega_{p}$ is $P_{sig} = \Delta \omega \cdot(k_{B}T/\pi \xi) /
(\omega_{c}^2 + \omega_{p}^2)$. The corresponding ratio of electric
to Brownian forces is $\gamma^{2} = \Delta \omega \cdot \omega_{c} /
\pi (\omega_{c}^2 + \omega_{p}^2)$ and the smallest detectable
particle charge is therefore,
\begin{equation}\label{eqn-detection-limit}
e|Z_{\rm{eff}}^{\rm{min}}| = \frac{\sqrt{12 k_{B}T \eta a \Delta
\omega}}{E}
\end{equation}
where we have used Eq.~\ref{eqn-Psignal}. In a discrete Fourier
transform the minimum width $\Delta \omega$ of a frequency channel
is determined by the total time $t_{0}$ of data acquisition and
equals, $\Delta \omega = 2 \pi / t_{0}$. In our system where $a =
610$ nm, $\eta = 1.38$ mPa s, and the maximum field strength is $E
\sim 80$ kV m$^{-1}$, the minimum detectable charge
 is about 0.25 $e$ for an experimental run of $t_{0}
= 26 $ s. This is equivalent to a limiting electrophoretic mobility
of $\sim 2.5 \times 10^{-12}$ m$^{2}$ V$^{-1}$ s$^{-1}$.

\subsection{Measurement of Charge Distributions}

We have shown that optical microelectrophoresis provides an
accurate, rapid and highly sensitive determination of the effective
charge on an individual trapped particle. In many cases, of course,
the population from which the single particle is selected is
heterogeneous. There is a distribution of particle charges which
reflects, for instance, the stochastic nature of the charging
mechanism or variations in the surface chemistry or the size of
individual particles. Many of the conventional methods used to probe
the electrokinetics of colloidal suspensions (microelectrophoresis,
laser Doppler electrophoresis, and electroacoustics) only allow for
the calculation of the average mobility. They provide either little
or no information on the distribution of particle charges. In many
cases, however, knowledge of the charge distribution is of vital
importance. Colloidal suspensions have, for example, proved to be
valuable model systems in condensed matter physics. There is strong
evidence that the width of the charge distribution (the charge
polydispersity) significantly influences the glass transition and at
high levels may suppress the freezing transition in these systems
\cite{4129}. It is therefore important that techniques are available
which allow an accurate characterization of the charge distribution.

We consider a suspension of colloidal particles with (effective)
charges $Ze$ which are distributed according to a normalized charge
distribution function $P(Z)$. The mean charge and charge
polydispersity $\sigma_{Z}$ are defined by
\begin{eqnarray}
  \overline{Z} &=& \int_{0}^{\infty} P(Z) Z \rm{d}Z  \\
  \sigma_{Z} &=& \frac{1}{\mid \overline{Z} \mid}
  \left ( \int_{0}^{\infty} P(Z) (Z- \overline{Z})^{2}
  \right )^{1/2}.
\end{eqnarray}
To characterize the charge distribution we trap and measure the
effective charge of 50--100 individual particles. Although
a little tedious these repeated measurements are still quick. For
instance it takes a couple of minutes per particle to manually
measure its charge: 20--30 s scanning time to locate a particle, a
short delay of 30 s to center the particle trap and a further 30 s
for data collection.

To explore the range of applicability and resolution of the
technique we determined the charge distribution in nonpolar
suspensions of PHSA-coated PMMA spheres. These colloidal spheres
have been widely used in fundamental studies of nucleation,
crystallization and glass formation since the coating with PHSA is
thought to provide a hard-sphere type interaction.
Figure~\ref{fig:exp-dist}(a) shows the charge distribution measured
for PHSA-coated PMMA spheres suspended in clean dry dodecane.
Clearly the particles are very weakly charged, with an average
colloidal charge of just three electrons per sphere, equivalent to
an electrophoretic mobility of $\mu = -3 \times 10^{-11}$ m$^{2}$
s$^{-1}$ V$^{-1}$ and a zeta potential of -3.5 mV. Indeed the
particle mobility was sufficiently small to be undetectable on a
commercial PALS instrument. The low mobilities however presented no
problems for the optical microelectrophoresis measurements, as the
data in Figure~\ref{fig:exp-dist}(a) confirms. Although the charge
on the PHSA-coated spheres is small, it is clearly non-zero and so
we expect a weak, soft repulsion between the particles. To gauge the
level of this repulsion we estimate the average contact value
$U_{0}$ of the interaction potential between spheres. An average
charge of $\overline{Z} = -2.9$ equates to a value of $U_{0} /
k_{B}T = 0.2$ in dodecane (Eq.~\ref{eqn-contact_pot}). Although the
value is small in comparison to thermal energies it can not be
ignored in quantitative studies. This is highlighted by, for
instance, the recent studies of Auer et al. \cite{4275} who have
explored the effect of a very similarly-sized weak repulsion on the
crystallization rate. They found that soft spheres crystallize some
two orders of magnitude faster than a comparable system of pure hard
spheres. Clearly, for quantitative studies, it is important to fully
characterize the interaction potential and optical
microelectrophoresis offers a promising alternative to existing
techniques.

The data in Figure~\ref{fig:exp-dist}(a) reveals a second
distinctive feature. The suspension is amphoteric, in that it
contains both positive and negatively-charged spheres. Close to
two-thirds of the particles are negative and the remainder positive.
While the charge distribution seems to be approximately Gaussian in
shape, the width is broad with a charge polydispersity of
$\sigma_{Z} = 1.5 \; \pm\; 0.3$, which is surprisingly large.

As a further example of the applicability of the optical
electrophoresis method, Figure~\ref{fig:exp-dist}(b) and (c) show
the effect of adding the surfactants,  Zr(Oct)$_{2}$ [b] and Na-AOT
[c], at concentrations of 2 mM to dilute PMMA suspensions in
dodecane. These surfactants form reverse micelles in nonpolar
solvents such as dodecane. A finite fraction of these micelles
spontaneously ionize at room temperature \cite{3312}. Particles then
acquire charge either by preferentially adsorbing the micelles of
one sign or by the dissociation of a surface group, with the
subsequent ionic species being solubilized within a reverse micelle.
While the micelle-mediated charging of colloids has been known for
at least the last 60 years \cite{Morrison-868}, the exact mechanism
of charging in nonpolar media remains problematic and is not well
understood. Figure~\ref{fig:exp-dist} reveals that addition of
Zr(Oct)$_{2}$ causes the PMMA spheres to become strongly positively
charged, while Na-AOT generates a large negative particle charge.
Using the H\"{u}ckel equation \cite{3479} we calculate the
equivalent zeta potentials to be +63 mV for the Zr(Oct)$_{2}$
treated spheres and -60 mV for the AOT containing suspensions. A
comparable value for the zeta potential of around -38 mV was found
by Kihara et al. \cite{3528} for PMMA particles in cyclohexane with
added AOT. While Hsu et al. \cite{3771} observed significantly
higher zeta potentials, of around -140 mV for PMMA in dodecane.  In
the case of Zr(Oct)$_{2}$, data is more limited, although Croucher
et al. \cite{3480} reports that poly(vinyl acetate) particles are
charged positive by the zirconyl salt. Surprisingly, we find that
the charge distribution of the zirconyl and AOT suspensions are very
sharply peaked with charge polydispersities of 5.0 $\pm$ 0.5\% for
the AOT system and 5.8 $\pm$ 0.6\% for ZrOEH. In both cases the
charge polydispersities are comparable, allowing for the sampling
errors, to the size polydispersity of the colloid (4.6 $\pm$ 1\%).
Detailed information on the charging mechanism in these systems will
be published elsewhere.

\section{Conclusions}

In summary, we have demonstrated a novel technique for the direct
measurement of small colloid charges commonly found in nonpolar
suspensions. The charge is measured by confining an individual
particle to a harmonic optical potential and by following the
excitation of the particle by a sinusoidal electric field. The
trapped particle forms a strongly damped oscillator. By using a weak
optical trap and an interferometric position detector we have shown
that single particle optical microelectrophoresis (SPOM) is capable
of very high sensitivity. Surface charges on the level of a few
elementary charges can be detected on individual colloidal
particles, with an uncertainty of about 0.25 e. Other techniques of
measuring the charge on a particle are generally less sensitive. Our
measurements are rapid and reproducible, with data on a single
particle being recorded in approximately 30 s. When applied to
multiple particles our technique yields information on the
distribution of particle charges rather than simply recording the
average charge, as provided by most conventional electrokinetic
techniques. These characteristics makes SPOM ideal for the
characterization of nonpolar suspensions and may prove useful to
study fundamental phenomena in colloid chemistry, such as the
relaxation of the double layer. For example,  one might be able to
detect discrete changes in the charge on a particle, due to the
time-dependent relaxation of the double layer. Work exploring these
possibilities is currently underway.

\begin{acknowledgments}
The authors wish to thank Unilever for partial funding and for
permission to publish these results. Support from the UK Engineering
and Physical Sciences Research Council is gratefully acknowledged.
Adele Donovan is thanked for help with particle synthesis and Kirsty
Paul for PALS measurements.

\end{acknowledgments}


\onecolumngrid

\section{Figures}

\begin{figure}[h]
\includegraphics[angle=0,width=7in]{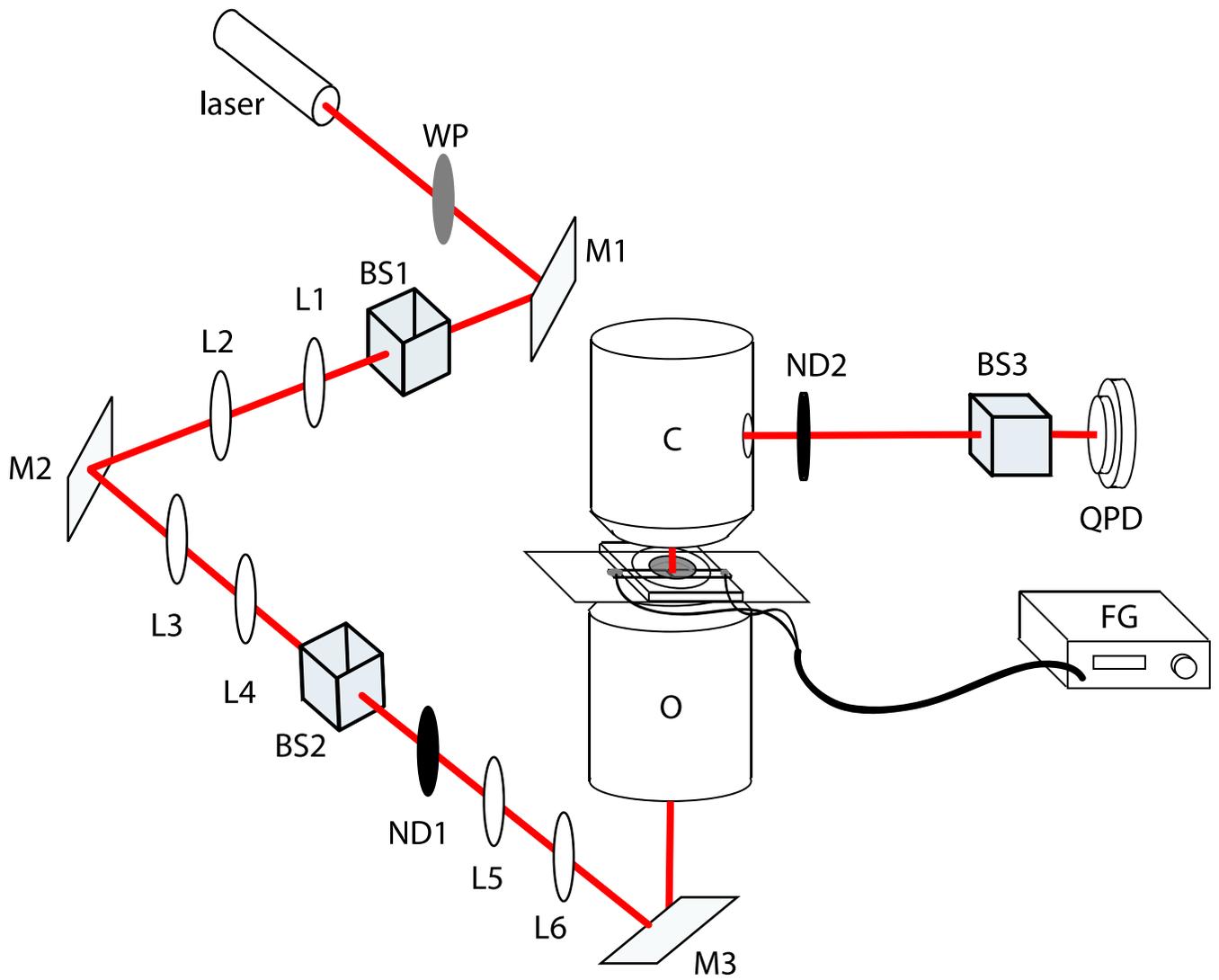}
\caption{Simplified experimental set-up of optical
microelectrophoresis apparatus. WP, $\lambda /2$-wave plate; M,
gimbal mirror; BS, beam-splitting cube; L, lens; ND, neutral density
filter; O, oil-immersion objective; C, high NA-condenser; QPD,
quadrant photodetector; FG, function generator.}
\label{fig:tweezer-schematic}
\end{figure}

\begin{figure}[h]
\includegraphics[angle=0,width=7in]{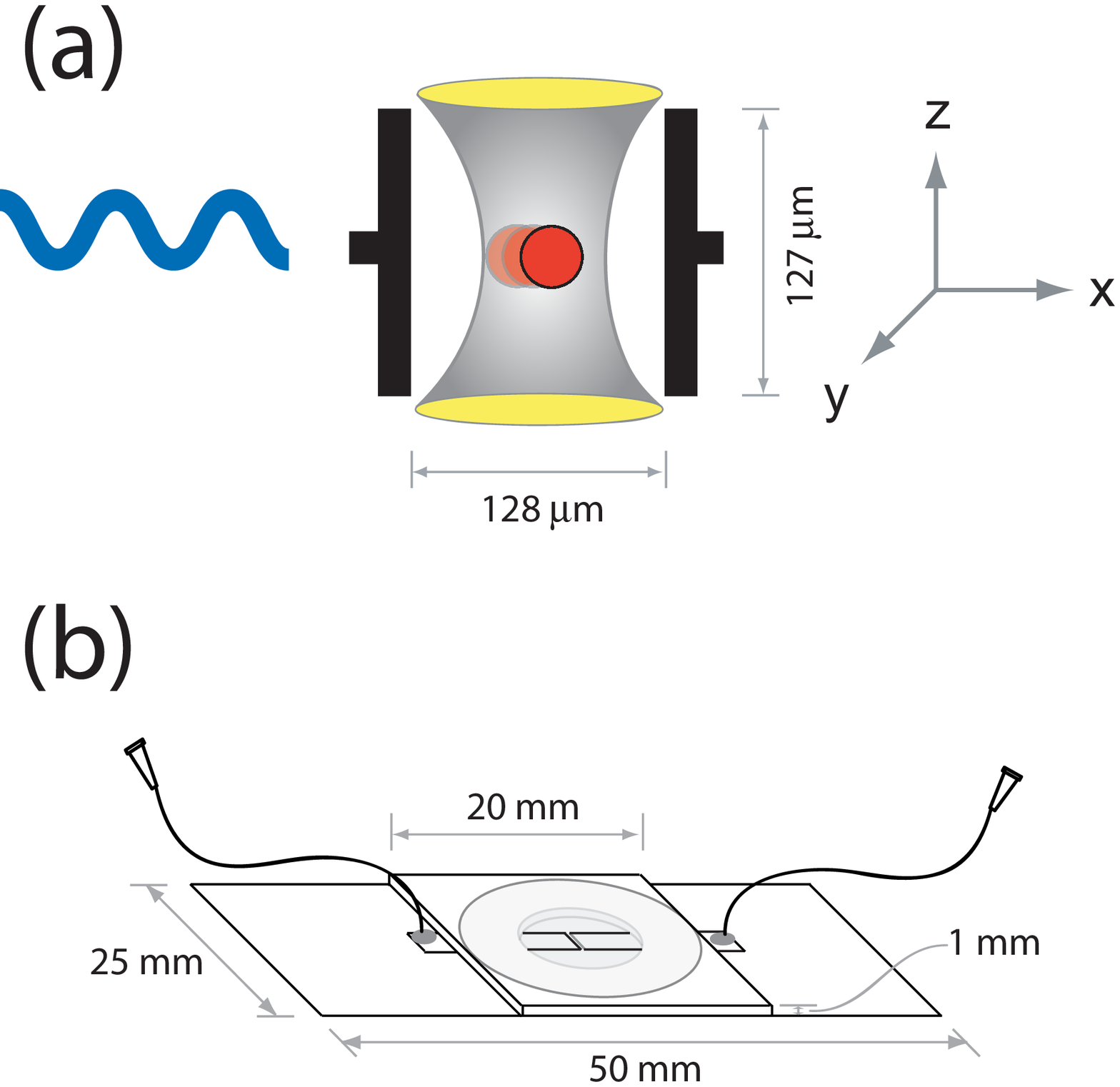}
\caption{(a) Experimental geometry for the optical
microelectrophoresis apparatus (not to scale). (b) Optical cell.}
\label{fig:tweezer-cell}
\end{figure}

\begin{figure}[h]
\includegraphics[width=7.0in]{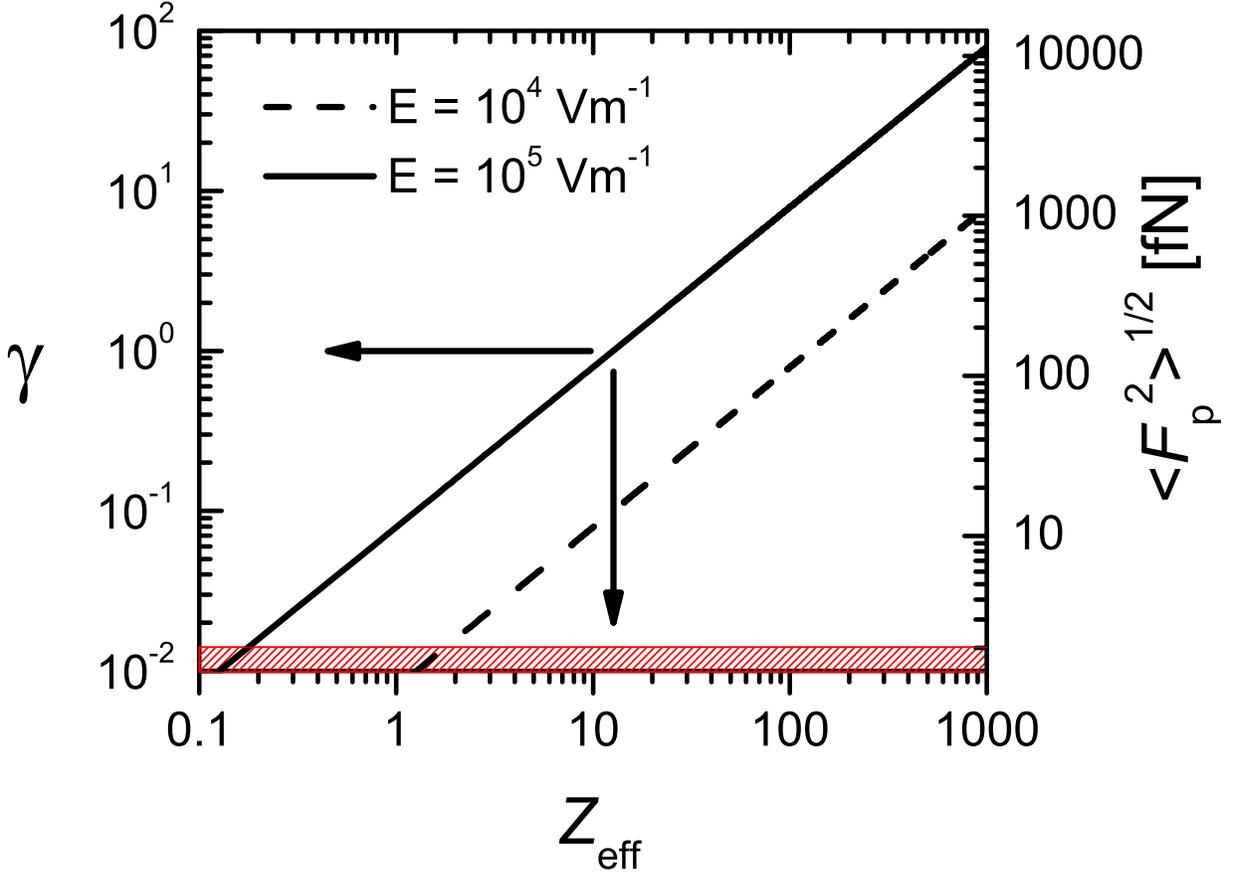}
\caption{ The dimensionless ratio $\gamma$ of electric and Brownian
forces. The trapped particle has a charge $eZ_{\rm{eff}}$ and is
placed in an oscillatory electric field of amplitude $E$ and
frequency $\omega_{p}$. The Brownian forces (shown on the right-hand
axis) are calculated assuming $\omega_{p} \ll \omega_{c}$ and a
force constant of $k_{H} = 5$ fN nm$^{-1}$. The shaded region
indicates the noise floor where the spectral density of the electric
and Brownian forces are equal at $\omega_{p}$. The boundary is
calculated assuming $a$ = 500 nm, $\eta = 1.38$ mPa s and $t_{0} =
26$ s. In the shaded region, the charge on the particle is too low
to be detectable by the current technique. The limiting charge
sensitivity is $\sim$ 2\;$e$ and 0.2 $e$ for fields of 10$^{4}$ and
10$^{5}$ Vm$^{-1}$, respectively.} \label{fig:curvature}
\end{figure}

\begin{figure}[h]
\begin{center}
\includegraphics[width=7.0in]{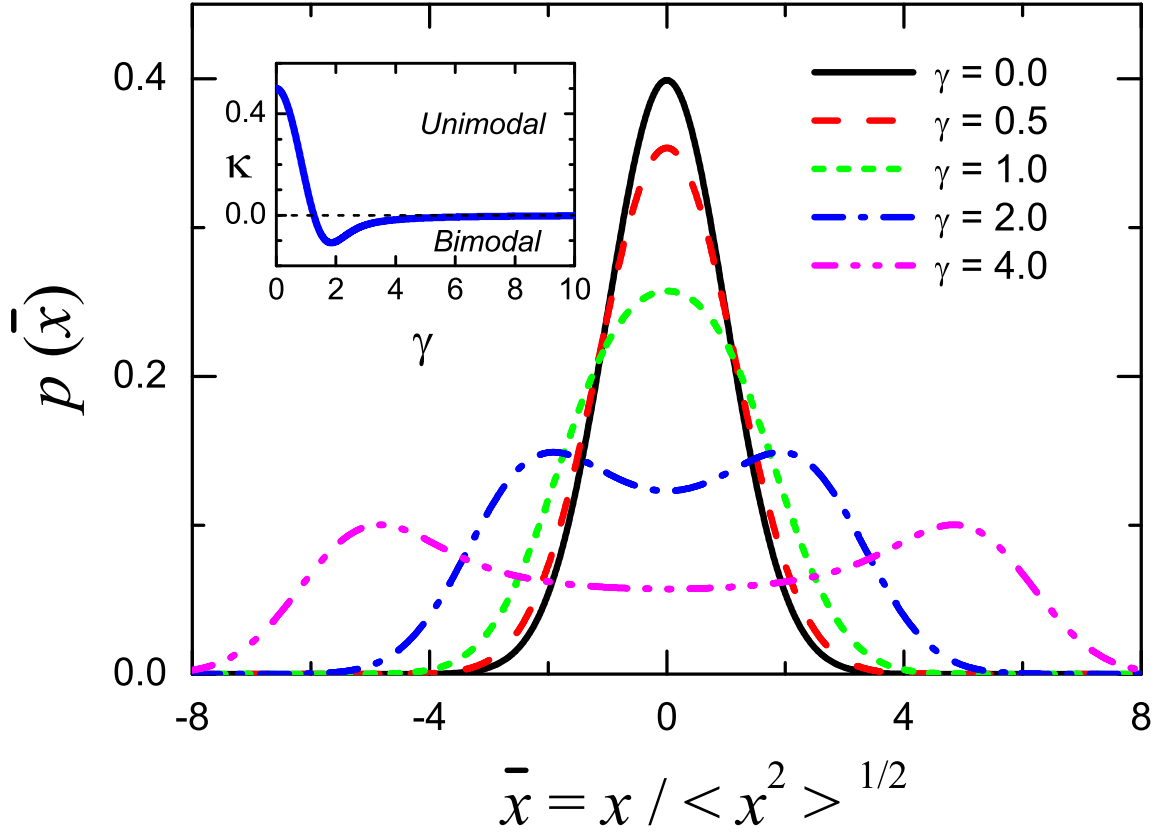}
\caption{The balance between thermal and electric forces. The full
lines show the dependence of the asymptotic probability
distribution, $p(\bar{x})$,  on the ratio $\gamma$ of the electric
and thermal forces. The inset plot shows the curvature $\kappa$ of
$p(\bar{x})$, evaluated at the origin $\bar{x} = 0$. The probability
distribution is double peaked for $\gamma \geq 1.257$ where $\kappa$
changes sign.} \label{fig:theoretical pdfs}
\end{center}
\end{figure}

\begin{figure}[h]
\includegraphics[angle=0,width=7in]{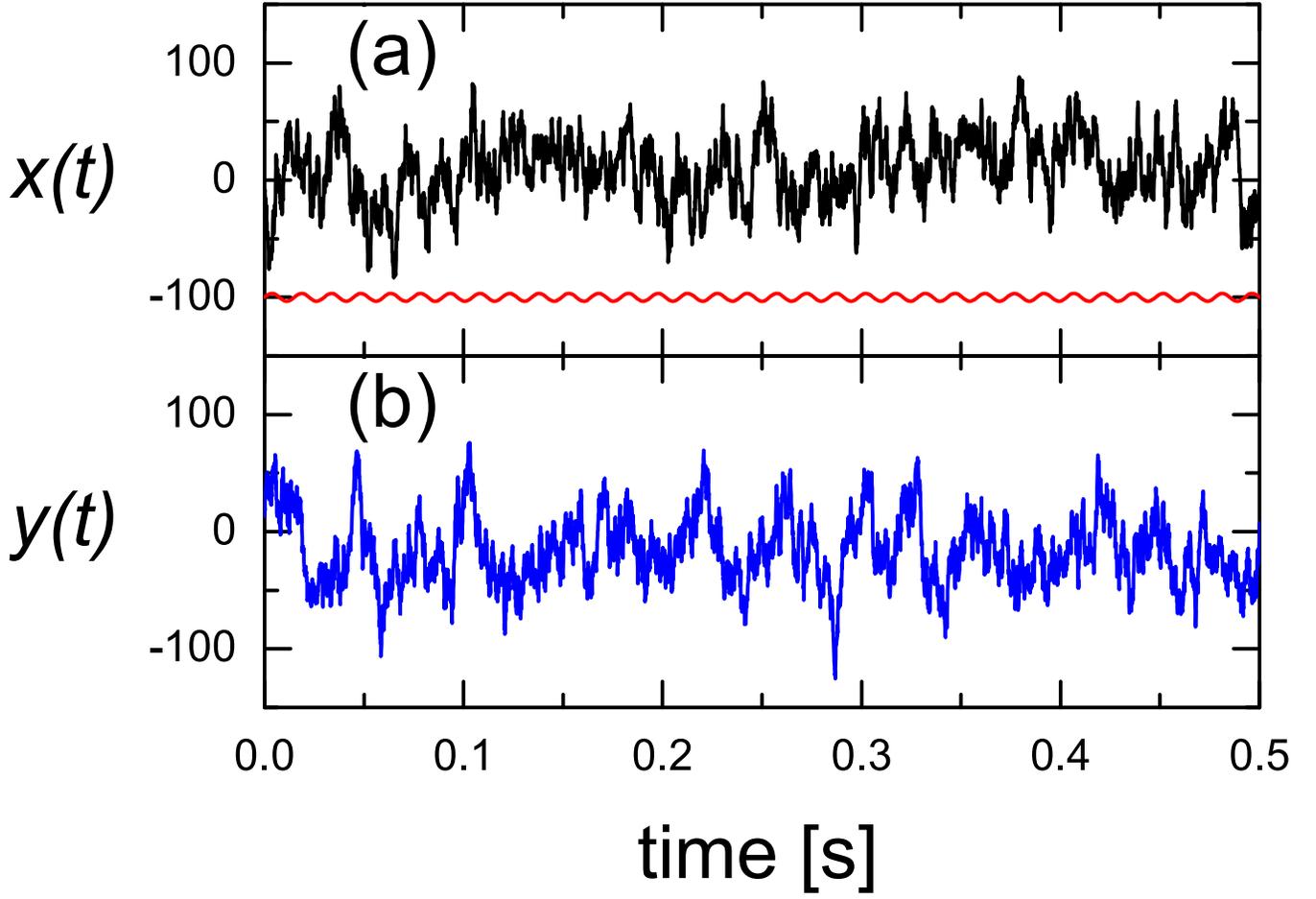}
\caption{Time trace of the $x$-- and $y$--coordinates of a charged
particle in a sinusoidal electric field. The field was aligned along
the $x$-axis and had a frequency of 67 Hz and an amplitude of 13.3
kV m$^{-1}$. Periodic oscillations are not visible in the measured
trajectory, which is dominated by random Brownian fluctuations. The
relative size of the periodic oscillations expected are illustrated
by the lower curve in (a).\label{fig:exp-trajectory}}
\end{figure}

\begin{figure}[h]
\includegraphics[angle=0,width=7.0in]{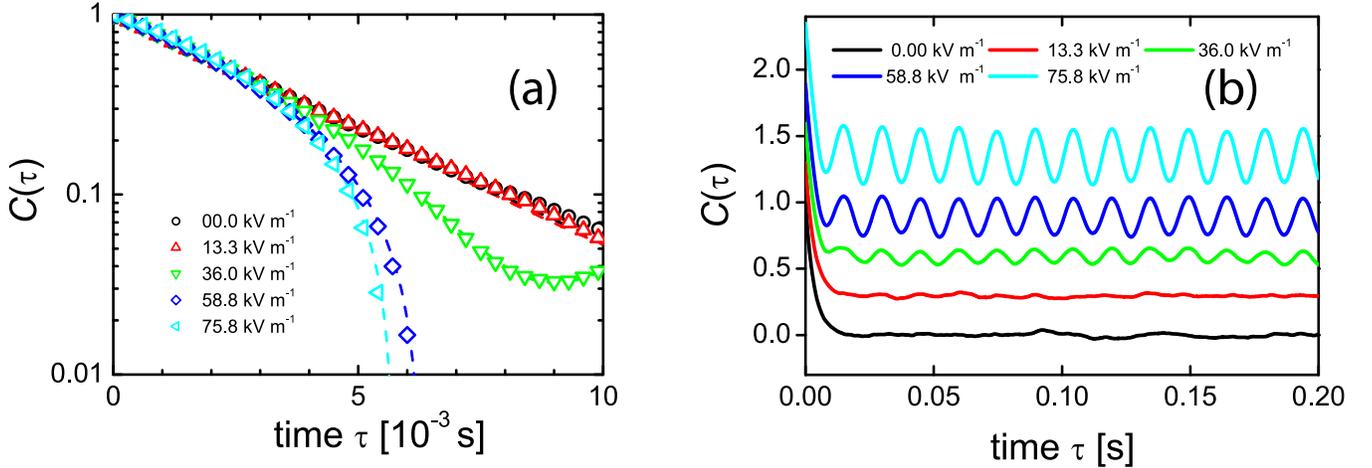}
\caption{The measured autocorrelation function $C(\tau)$. The sample
was a PMMA particle in dodecane (containing 0.035 wt\% free
PHSA-g-PMMA copolymer). Data was recorded for 26 seconds at each
field. The driving frequency $\omega_{p}$ was 420.8 rad s$^{-1}$.
Fig.~(a) shows that at short delay times the measured
autocorrelations collapse to a single exponential decay. The dashed
lines in (a) are least-squares fits to Eq.~\ref{eqn-norm-correlate},
with a corner frequency of 295 $\pm$ 7 rad s$^{-1}$. Fig.~(b)
reveals the periodic correlations found at long delay times. The
curves have been vertically displaced for clarity. }
\label{fig:autocorrelations}
\end{figure}

\begin{figure}[h]
\includegraphics[angle=0,width=7in]{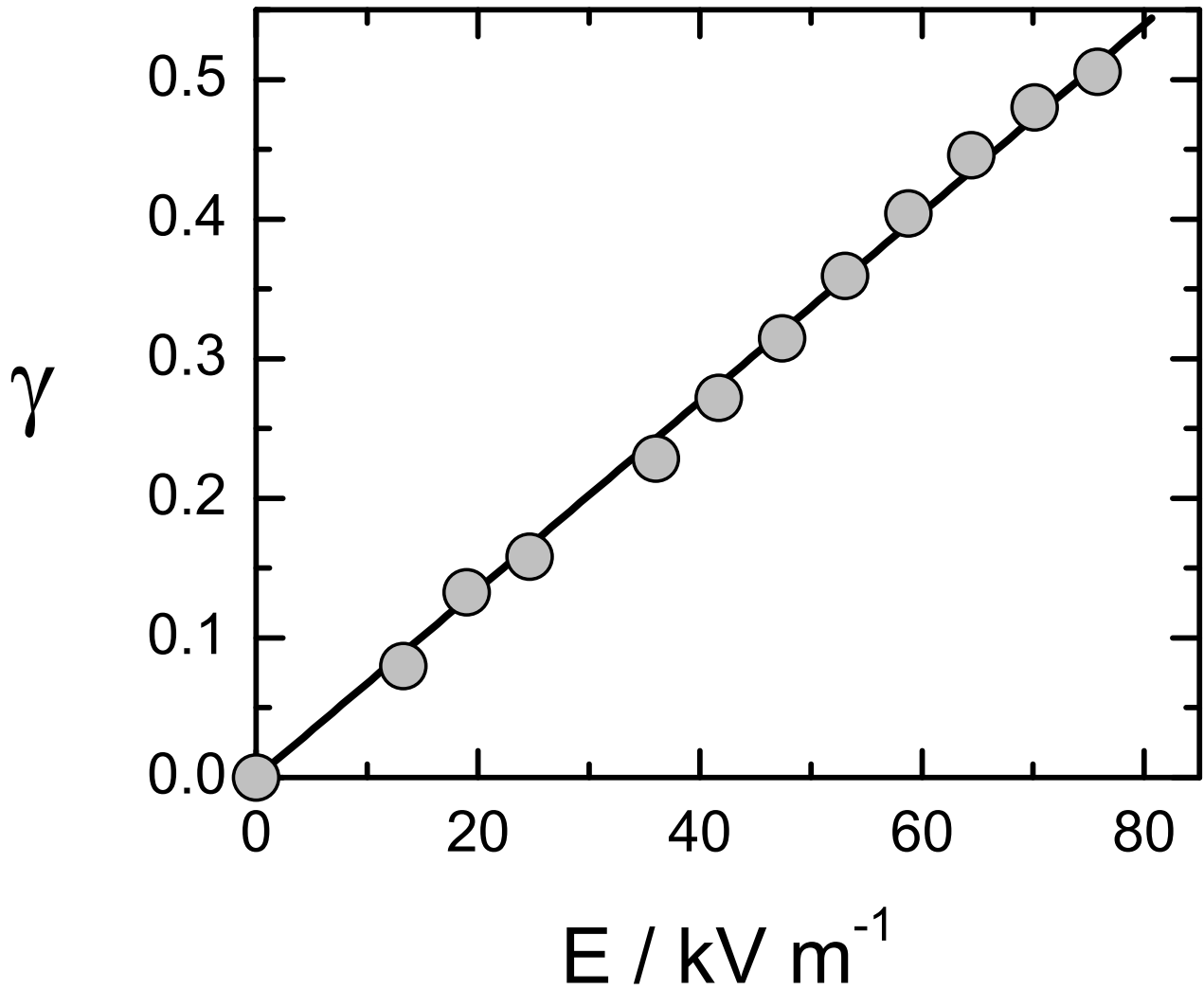}
\caption{The field dependence of the force ratio $\gamma$. $\gamma$
was obtained by fitting the autocorrelations displayed in
Fig.~\ref{fig:autocorrelations} to Eq.~\ref{eqn-norm-correlate}. The
linear variation of $\gamma$ with the field $E$ confirms that the
charge on the particle is independent of the strength of the applied
field. The solid line yields an effective charge of $|Z_{\rm{eff}}|
= 14.5 \pm 0.2$.} \label{fig:field-dependence}
\end{figure}

\begin{figure}[h]
\includegraphics[angle=0,width=7in]{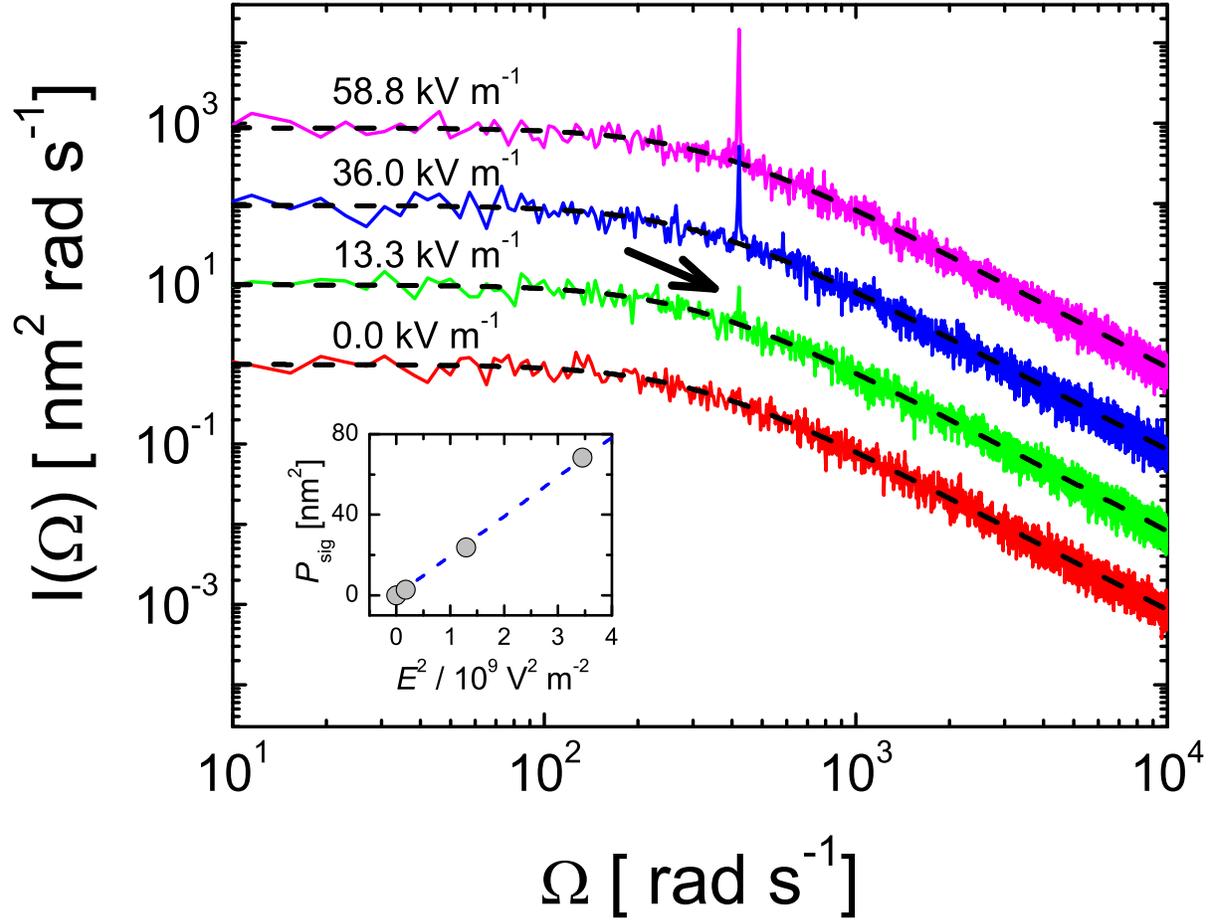}
\caption{The field dependence of the spectral density $I(\Omega)$.
The sample was a 610 nm PMMA particle suspended in dodecane with
0.035 wt\% added PHSA-g-PMMA copolymer. The variance of the power
spectrum was reduced by averaging together 64 spectra, calculated
from consecutive trajectories each containing $2^{14}$ data
points ($t_{0} = 1.64$ s) recorded on the same particle.}
\label{fig:exp-power}
\end{figure}

\begin{figure}[h]
\includegraphics[angle=0,width=7in]{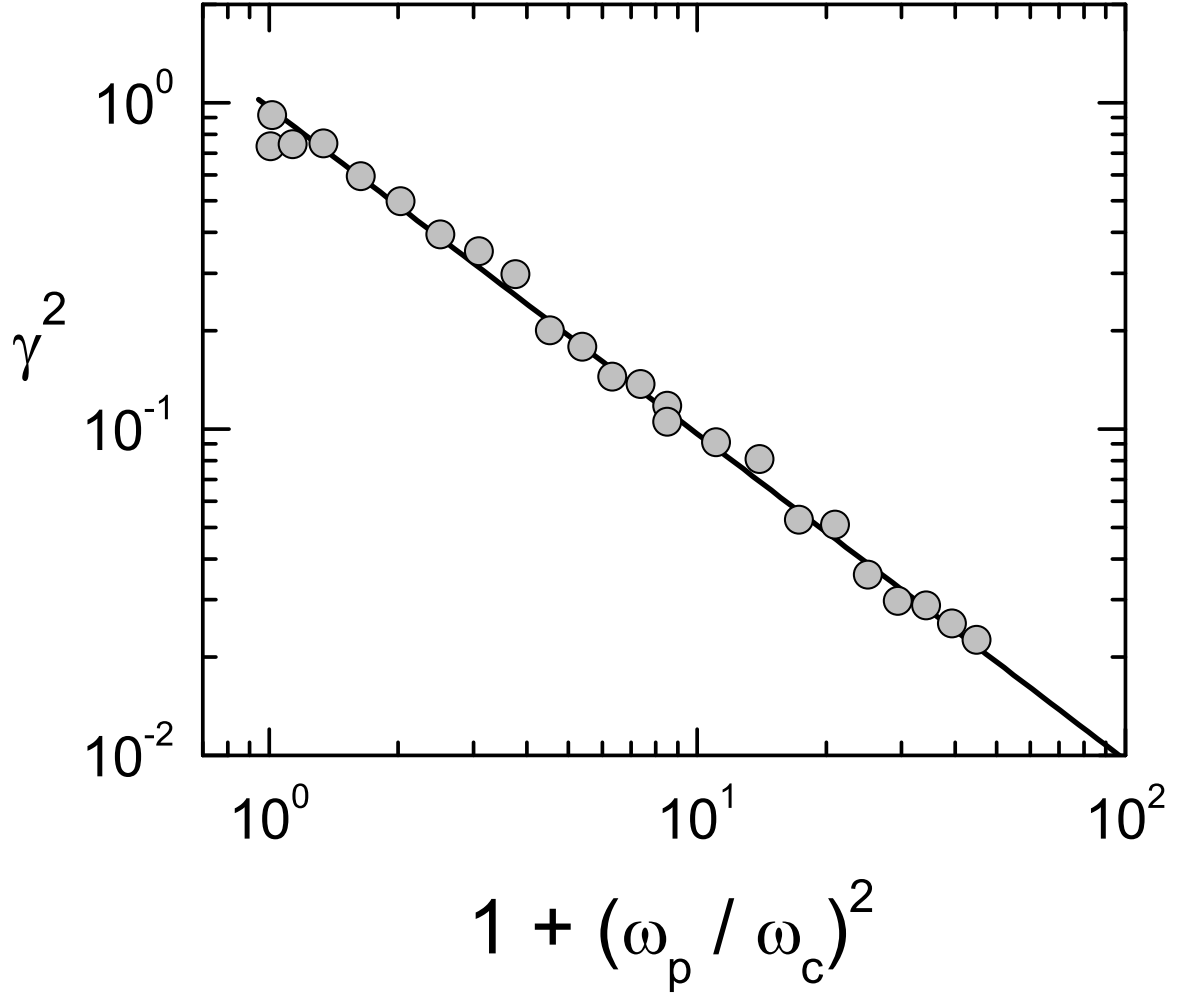}
\caption{The frequency dependence of the force ratio $\gamma$. The
amplitude of the electric field was fixed at $E = 75.8$ kV
m$^{-1}$ and the frequency $\omega_{p}$ varied.  The solid line is a
least-squares fit of $\log \gamma^{2}$ against $\log (1 +
(\omega_{p}/\omega_{c})^{2})$ with a gradient of -0.98 $\pm$ 0.02.
\label{fig:exp-freq}}
\end{figure}

\begin{figure}[h]
\includegraphics[angle=0,width=7in]{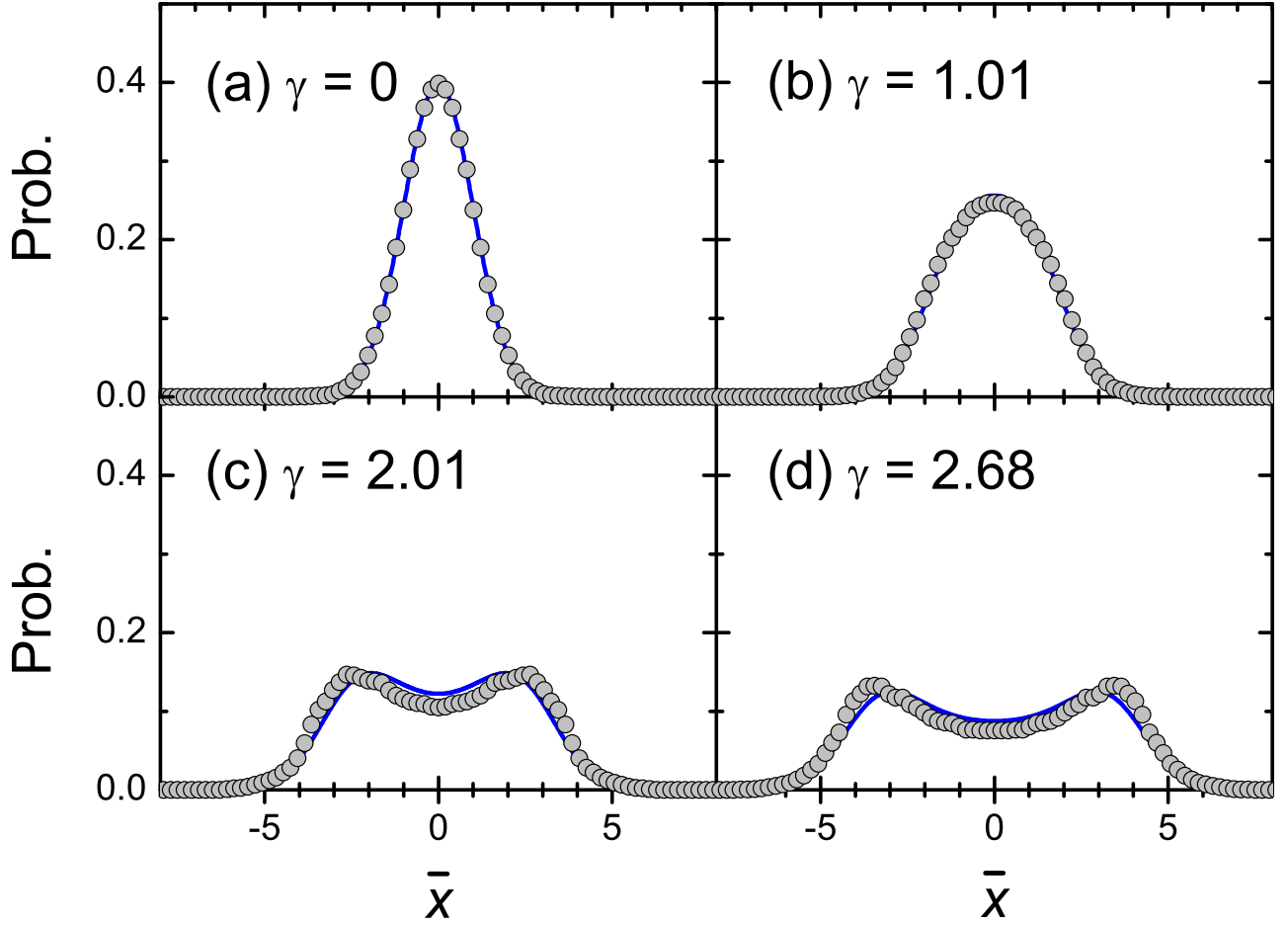}
\caption{The probability distributions characterising the restricted
diffusion of an optically-trapped PMMA particle in a solution of 100
mM Na-AOT in dodecane. The corner frequency of the optical trap was
25.0 Hz. The points give the measured distribution in electric
fields of amplitude (a) 0 kV m$^{-1}$, (b) 22.8 kV m$^{-1}$, (c)
45.5 kV m$^{-1}$, and (d) 60.7 kV m$^{-1}$. The sinusoidal electric
field had a frequency of 17.5 Hz. The solid lines were calculated
from the theory for a driven harmonically-bound particle
(Eq.~\ref{eqn-general-pdf}) using experimentally-measured
parameters. The comparison has no adjustable parameters.}
\label{fig:exp-pdf}
\end{figure}

\begin{figure}[h]
\includegraphics[angle=0,width=7in]{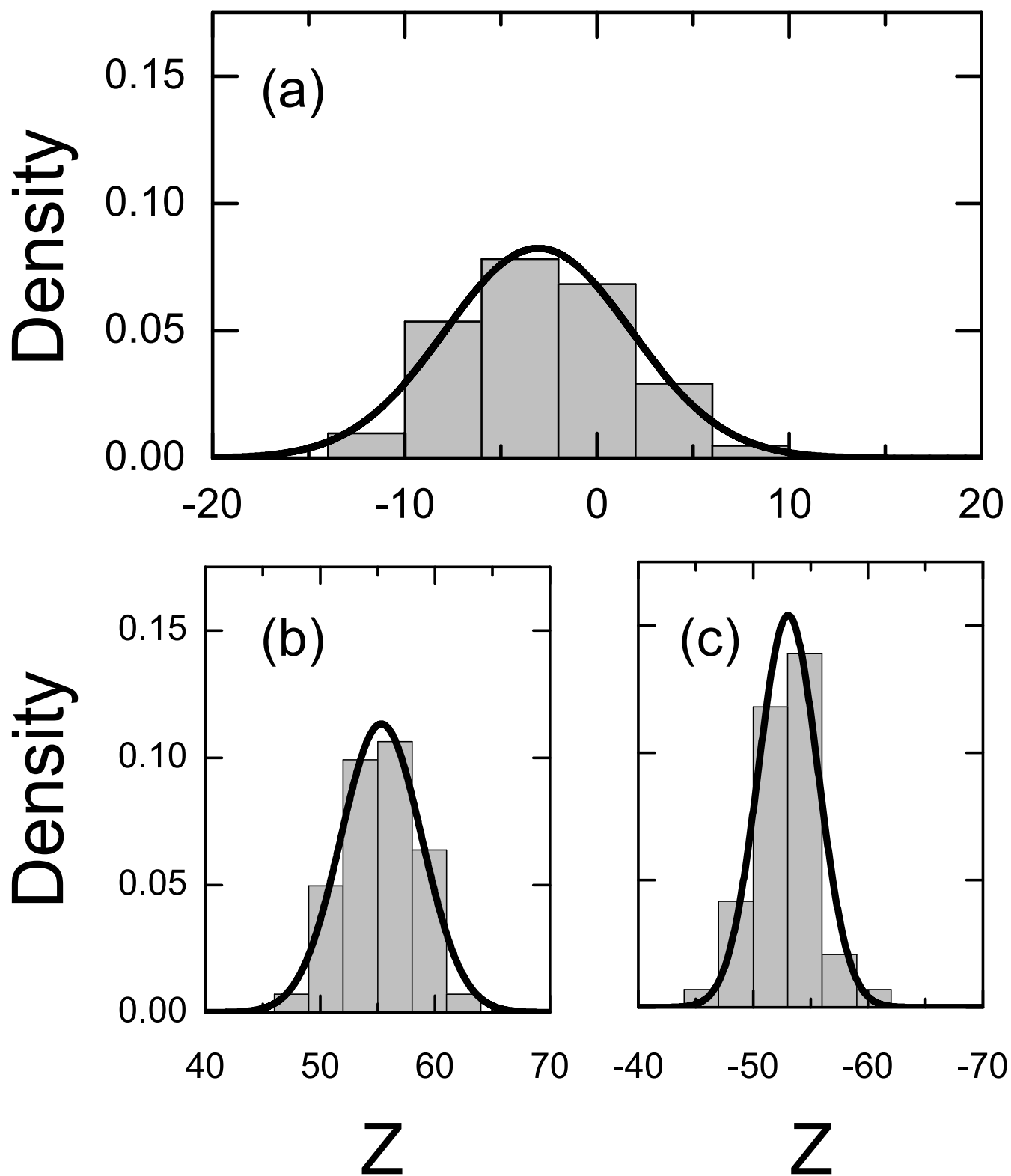}
\caption{The charge probability distribution measured for PMMA
particles in dry dodecane with (a) no added surfactant, (b) 2 mM
zirconyl 2-ethyl hexanoate, and (c) 2 mM sodium-AOT added. The full
lines are Gaussian fits. \label{fig:exp-dist}}
\end{figure}

\end{document}